\documentclass[12pt,a4paper]{article}

\usepackage{pslatex}
\usepackage{graphicx}
\usepackage{setspace}
\usepackage{array}

\usepackage[ps2pdf, colorlinks, bookmarks, pdfview={FitH}, pdfstartview={FitH}]{hyperref}

\def\noi{\noindent}

\begin{document}

\date{}

\title{{\bf Avoiding Chaos in Wonderland}\\
       }

\author{G.A. Kohring\\
        C\&C Research Laboratories, NEC Europe Ltd. \\
        Rathausallee 10, D-53757 St. Augustin, Germany\\
        kohring@ccrl-nece.de
}

\maketitle

\bigskip

\begin{abstract}
Wonderland, a compact, integrated economic, demographic and environmental 
model is
investigated using methods developed for studying critical phenomena.
Simulation results show the parameter space separates into two phases, one of
which contains the property of long term, sustainable development.
By employing information contain in the phase diagram, an optimal strategy 
involving pollution taxes is developed as a means of moving a system initially
in a unsustainable region of the phase diagram into a region of sustainability
while ensuring minimal regret with respect to long term economic growth.
\end{abstract}

\bigskip
\noindent{\bf PACS-2003:}
89.65-s,89.65.Gh,89.75.-k

\noindent{\bf Keywords:}
Wonderland Model, Critical Phenomena, Complex Systems, Sustainable Development

\bigskip

\noindent Submitted to \textit{Physica A}.

\newpage

\section{Introduction}

Long-term planning to ensure a sustainable world is a difficult problem for
many reasons, not the least of which is the lack of certainty surrounding
mankind's interactions with his environment. Given a host
of models with varying levels of complexity \cite{CHA00}; it is difficult
to develop a unified methodology for understanding the intricate dynamics
contained within these models.  A common 
approach to problem of long-term planning is the \textit{Scenario} method 
\cite{MRM04,HUG99,RBG02}; whereby, the goal is to create a small set of 
reasonable scenarios extrapolating from the present into possible futures.
These scenarios are often supported by quite large computer simulations
of models
describing the different sectors of an integrated world. Generally, these 
models
include economic, demographic and environmental sectors \cite{MRM04}, though 
they may also include political \cite{HUG99} and security \cite{RBG02}
components as well. Basically, the idea behind the scenario approach is to 
selects a reasonable set of
parameters for the model in question, then integrate the model forward over
the planning time frame in order to ascertain what type of world might evolve.

While the scenario approach yields readily understandable projections based upon
the current state and justifiable assumptions; its basic weakness is the
complexity of the models themselves. The values of some of the parameters which
enter the models are not know with any accuracy and can only be inferred from 
the current state of the system; at the same time
some of the equations describing the system are chaotic. This combination of 
vaguely known
parameters and chaotic equations of motion cast doubt on the reliability of
the forecasts \cite{CAM05}, leading critics to charge the authors with
cherry picking, i.e., selecting sets of parameters to fit the authors 
preconceived notions of how the world should evolve.

Recently, Bankes, et al. \cite{BLP01,LPB03} suggested a more fundamental 
approach
based upon large scale computer simulations, specifically aimed at avoiding
the charges of cherry picking. By using Monte Carlo techniques to create a
large ensemble of scenarios, they propose to avoid bias towards any one 
scenario. Projections used for long-term planning are then based upon the 
probability of any given scenario occurring in the ensemble.

In this paper the ensemble approach is extended to a systematic exploration
of the parameter space, whereby the equations of motion are treated as 
equations of state, leading to a multi-dimensional phase diagram. The phase
diagram allows a clean separation of the parameter space into regions
where sustainable development is possible and where it is not. This removes
the guess work from the construction of plausible scenarios and brings more 
focus to the ensemble method.

For discussing the current proposal in more depth, the Wonderland model 
\cite{SAN94,HL98,LPB03}\nocite{LUT94} has been chosen as an example
because of its relative simplicity and tractability. Although it lacks the
details of many other models, it still captures enough general behavior
of more sophisticated models to make its study worthwhile.

The next section describes the Wonderland model in more detail, after which
a detailed discussion of Wonderland's behavior and phase diagram is
undertaken. Based upon these findings, section \ref{sec:tax} presents a
strategy for long term planning aimed at ensuring sustainable development.

\section{The Wonderland Model}

Sanderson's Wonderland model \cite{SAN94,HL98}\nocite{LUT94} 
describes in an integrative 
framework the economic, demographic and environmental sectors of an idealized
world. The model is characterized by four state variables, $\{x(t),\ y(t),\ 
z(t),\ p(t)\}$, representing the population, per capita output, stock of 
natural
capital and pollution flow per unit of output respectively. These four state 
variables evolve according to following set of non-linear, difference
equations:

\begin{eqnarray}
x(t+1) &=& x(t)\left[1+n\Big(y(t),z(t)\Big)\right],
\label{eq:x_state} \\
y(t+1) &=&
y(t)\left(1+\gamma-(\gamma+\eta)\Big[1-z(t)\Big]^{\lambda}\right),
\label{eq:y_state}\\
z(t+1)
&=&\frac{g\Big(x(t),y(t),z(t),p(t)\Big)}{1+g\Big(x(t),y(t),z(t),p(t)\Big)},
\label{eq:z_state}\\
p(t+1) &=&p(t)(1-\chi).
\label{eq:p_state}
\end{eqnarray}

\noi The state variables for population ($x$) and per capita output ($y$) can 
assume all non-negative real
values ($x,y\in[0,\infty)$), while the stock of natural capital ($z$) and the
pollution per unit of output ($p$) are confined to the unit 
interval ($z,p\in[0,1]$).  A value of $z=1$ represents a full stock of
unpolluted natural
capital and $z=0$ represents the fully polluted state. $p=1$
on the other hand represents maximal pollution per unit of output and $p=0$
implies no pollution per unit of output.

In eq.~\ref{eq:x_state},
the endogenous population growth rate, $n(y,z)$, can be written as
as the difference between the crude birth rate, $b(y,z)$ (number of births
per $1\,000$ population per time at a given time step) and the crude
death rate, $d(y,z)$ (number of deaths per $1\,000$ population per time at a
given time step):

\begin{eqnarray}
n(y,z)&=&b(y,z) - d(y,z)
\label{eq:n} \\
b(y,z)&=&\beta_0\left[\beta_1 - \left(
         \frac{e^{\beta y}}{1+e^{\beta y}}\right)\right],
\label{eq:b} \\
d(y,z)&=&\alpha_0\left[\alpha_1 - \left(
         \frac{e^{\alpha y}}{1+e^{\alpha y}}\right)\right]
         \left[1+\alpha_2(1-z)^{\theta}\right],
\label{eq:d}
\end{eqnarray}

\noi The parameters $\beta$, $\beta_0$ and $\beta_1$ govern the birth rate,
while the parameters $\alpha$, $\alpha_0$, $\alpha_1$, $\alpha_2$ and
$\theta$ govern the death rate.
From eqs.~\ref{eq:b} and \ref{eq:d} it can be 
seen how both the birth rate and death rate decrease with increases in the
per capita output, $y$. Furthermore, in eq.~\ref{eq:d}, the death rate 
is seen to increase when the environment deteriorates, i.e., when $z$ decreases.
These effects are in line with recent studies relating population growth 
with industrial output \cite{COH95,BLE05}.

The function $g(x,y,z,p)$ which determines the time evolution of the
natural capital is given by:

\begin{eqnarray}
g(x,y,z,p)&=& \frac{z}{1-z}\,e^{\,\delta z^{\rho}-\omega f(x,y,p)
                 },
\label{eq:g} \\
\end{eqnarray}

\noi where, $f(x,y,p)$, is the pollution flow:

\begin{equation}
f(x,y,p)=xyp.
\label{eq:f}
\end{equation}
The parameters $\delta$, $\rho$, and $\omega$ determine the pollution
flow at which the economic, environmental and demographic sectors are in
balance.  This critical pollution flow,
$\delta z^{\rho}/\omega$, determines the rate at which the natural capital is
able to ameliorate the pollution flow, $f$. As can be seen from 
eqs.~\ref{eq:z_state} and \ref{eq:g}, when $f=\delta z^{\rho}/\omega$
the level of natural capital remains constant and the economic sector is in
balance with the environmental sector.

The form of $f$ originates from the I-PAT hypothesis \cite{EH71}. In its
original form, the I-PAT hypothesis states: a
population's impact on its environment is equal to its size multiplied by its 
per capita
output and its level of technology. In wonderland, the impact is the
pollution flow and the level of technology is
represented by the pollution per unit of output; hence, $f=xyp$.

Wonderland's economic sector is characterized by the parameters:
$\gamma$, $\eta$, $\lambda$ and $\chi$. $\gamma$ is the exogenous economic 
growth
rate and determines how fast the economy could grow if its capital stock
were fully intact (see eq.~\ref{eq:y_state}). $\eta$ and $\lambda$ determine 
how rapidly the economy deteriorates (recovers) when the capital stock declines
(increases).  The parameter $\chi$ governs the economic decoupling rate, 
i.e., the rate at
which technological innovations reduce the pollution flow per unit of output.

Some variants of the Wonderland model include a term in eq.~\ref{eq:f}
governing pollution control expenditures \cite{GWM96,MPF96,HL98,LH02}.  
Since pollution
control is essentially a policy decision and not an intrinsic part of the
model, in this paper, we
follow \cite{LPB03} and introduce pollution control later
in section~\ref{sec:tax} where long term planning is discussed in more
detail. Such an
approach allows us to cleanly separate the intrinsic dynamics of the
model as given above, and perturbations of the dynamics due to policy
decisions.

As can be seen, equations~\ref{eq:x_state} to \ref{eq:f} depend upon 
15 positive parameters which govern the overall behavior of model; whereby
eight of them ($\alpha$, $\alpha_0$, $\alpha_1$, $\alpha_2$,
$\beta$, $\beta_0$, $\beta_1$ and $\theta$) determine the population growth,
four ($\delta$, $\rho$, $\omega$ and $\chi$) determine the state of the
environment and three ($\gamma$, $\eta$ and $\lambda$) determine the health
of the economy. In the next section we take some slices through this 15
dimensional parameter space to learn more about the models behavior.

\section{Phase Diagram \label{sec:phase}}

As noted in the introduction, previous work on this model has mostly followed
a scenario based approach \cite{SAN94,GWM96,BLP01}, with the two most often
investigated scenarios being the so-called ``Dream'' scenario and the ``Horror"
scenario. The Dream scenario earns its name because it holds out the
possibility of continued economic growth combined with a stable population and 
a healthy environment; whereas the Horror scenario depicts environmental
collapse followed by an economic collapse and a declining population.  
Table~\ref{tab:dream} contains the parameters for the Dream scenario. In the
horror scenario, the only parameter whose value is changed, is the
decoupling rate, $\chi$, which takes on the new value $\chi=0.01$.

Plots of the time evolution for the Dream and Horror scenarios are shown in 
Figure~\ref{fig:traj}. Initially, both scenarios follow the same growth curves
in terms of per capita output and population; however, after 
approximately 90
years, the world in the Horror scenario undergoes a spontaneous transition to
a phase marked by a depleted stock of natural capital,
an economic depression and population decline.  The sudden collapse of the 
environment after 90 years of
relative stability is indicative of a first order phase transition.

Until now, our analysis has concentrated upon the classical scenario approach.
As a first step in going beyond this approach, we examine the behavior of the
model when systematically moving through the parameter space along the line
$\chi=0.01$ to $\chi=0.04$ while holding all the other parameters fixed. 
For this
purpose, define an order parameter, $t_c$, as the number of years before the
natural stock collapses. The justification of calling $t_c$
an order parameter \cite{SMA76} becomes apparent when examining 
Figure~\ref{fig:scale}. As can be seen, $t_c$ follows the scaling
behavior expected in the presence of a second order phase transition:

\begin{equation}
t_c\sim \Big(\chi -\chi_c\Big)^{\zeta},
\label{eq:tc}
\end{equation}

\noi furthermore, $t_c$ is undefined for all $\chi \geq \chi_c$. From the data
in Figure~\ref{fig:scale}, the critical value of $\chi$ can be 
estimated: $\chi_c\approx 0.0385$, along with the critical exponent,
$\zeta~\approx~0.945$. The symbols in Figure~\ref{fig:scale}, indicate different
starting states.  Depending upon the exact initial state, the collapse may 
take place sooner or latter, but the scaling behavior is the same. From this
data we can view $\chi_c$ as marking a phase boundary between the phase
of unsustainable development,
as epitomized by the Horror scenario, and the phase of sustainable development
as epitomized by the Dream scenario. (The reader should not take this diagram
to mean we are advocating a planning horizon extending to $100\, 000$ years!
Rather, the diagram indicates how rapidly $t_c$ changes given small changes in
$\chi$.)

In the Wonderland model, the parameters most responsible for the
interactions of the economic and environmental sectors are $\gamma$ and
$\chi$. Therefore, if we repeat the above analysis for the remainder of
the $\gamma-\chi$
plane, we arrive at the phase diagram shown in Figure~\ref{fig:phase}.
The line of points delineates the parameter space into a phase of sustainable
development and one of unsustainable development. In the phase of sustainable
development the economic, demographic and environmental sectors of Wonderland
are in equilibrium, while in the phase of unsustainable development, these
sectors are out of equilibrium, eventually leading to a world wide collapse. 
As one
approaches the phase boundary from below, the time before the collapse occurs
increases as a power law of the distance from the boundary.

One can continue this exercise for the remaining parameters in the model;
however, most of the other parameters effect the quantitative results but not
the qualitative features described above.  The parameters $\delta$, $\rho$ and
$\omega$, for example, determine the exact value of the critical 
pollution flow above which 
the environment begins deteriorating, but not the existence of a critical
pollution flow \cite{GWM96}.

\section{Policy Planning \label{sec:tax}}

In the previous section, the analysis focused on the intrinsic behavior of the
Wonderland model; however, this is only part of the problem, the other more
intricate question is whether or not it is possible to introduce control
mechanisms in order to avoid the phase transition in the Horror scenario and
the  attendant catastrophic consequences.
Note, the model as described by
equations \ref{eq:x_state} through \ref{eq:f} has no steerable parameters. All
of the 15 parameters entering into the Model are in principal measurable
or can be estimated using available data \cite{SAN94,LPB03}; hence, once they
are known one can look on the phase diagram of Figure~\ref{fig:phase} to 
determine whether the world of Wonderland is in a sustainable phase or
an unsustainable phase approaching a potentially catastrophic collapse.

When faced with the question of how to handle sustainable development,
policy makers can choose to either control the emission of pollutants at their
source or to spend funds abating pollutants aftwards.  Either of these 
approaches may be effective or
they may induce undesirable side affects. As a first step, we look at the
the pollution abatement approach.

\subsection{Pollution Abatement}

To abate the effects of pollution, funds must be drawn from other sources
of income; furthermore, as the environmental degradation becomes more serious,
more funds are required. A simple, non-linear model governing the
expenditures for pollution abatement, 
$c(y,z)$, has been proposed by previous authors \cite{SAN94,HL98,LH02}:

\begin{equation}
c(y,z)=\phi(1-z)^{\mu}y.
\label{eq:c}
\end{equation}
In this model, as natural capital deteriorates, the expenditures to abate 
pollution
increase, whereby the rate at which expenditures increase are governed by the
policy parameters $\phi$ and $\mu$. Withdrawing capital in this manner,
decreases the per capita output available for other uses; hence, 
the per capita output, $y$, in equations
\ref{eq:b} and \ref{eq:d} must be replaced by $y^{\prime}=y-c$.
(Eq.~\ref{eq:y_state}
does not change, because goods and services needed for pollution abatement 
are part of the overall output.) Furthermore,
since the aim is to improve the state of the environment, the pollution 
flow is reduced by the effectiveness of the regulations, i.e., 
eq.~\ref{eq:f}, becomes\cite{SAN94,HL98,LH02}:

\begin{equation}
f(x,y,p)=xyp - \kappa\frac{e^{\epsilon cx}}{1+e^{\epsilon cx}}.
\label{eq:freg}
\end{equation}
where, $\kappa$ determines the effectiveness of the expenditures, $c(y,z)$.

Basically, policy makers have the parameters $\phi$, $\mu$ and $\kappa$ 
for control purposes, though none of these parameters can be varied without
limits. $\phi$ and $\mu$ determine how much output is diverted to pollution
abatement once the stock of natural capital starts deteriorating. Since we 
require $c<y$, this limits $\phi < 1$. $\mu$ determines how quickly the
inhabitants of Wonderland respond to early signs of environmental degradation.
While these parameters cannot of themselves drive the system from the phase
of unsustainable development to the phase of sustainable development, 
they do have an impact on the amount of chaos present in the 
unsustainable phase.

An intriguing new phenomena in this model is the environments recovery 
from collapse. Indeed, one can simulate the Horror scenario over several 
millennium,and observe how it undergoes repeating cycles of
collapse and recovery. To gain insight into the underlying dynamics it is
instructive to plot the orbits of the real growth rate as a function of
the stock of natural capital. The normalized real grow rate can be defined as:

\begin{equation}
r(t)=\frac{y(t)-y(t-1)}{\gamma\, y(t-1)},
\label{eq:r}
\end{equation}

\noi where we have normalized the real growth rate by the exogenous growth
rate, $\gamma$. A plot of $r(t)$ versus $z(t)$ is shown in
Figure~\ref{fig:chaos}.  When the environment is deteriorating, the economy
follows the upper curve and when the environment is improving, the economy
follows the lower curve.  Hysteresis curves of this type are expected in the
presence of a first order phase transition.
From this figure it can be seen that the momentary
performance of the economy sheds little light on the overall health of
Wonderland, since the growth rate at first decreases only slowly with
the declining capital stock until the natural capital is nearly exhausted, at
which point the growth rate rapidly turns from positive to negative. The exact
shape of the orbits depend upon the parameters $\eta$ and $\lambda$ from 
eq.~\ref{eq:y_state}.  This
general picture is consistent with scenarios described by more detailed models
such as World3 \cite{MRM04}, thereby lending credence to the proposition of 
using Wonderland as a tractable model for in depth studies.

Notice how
the trajectories of the Horror scenario appear to contract to an aperiodic
recurrent attractor \cite{CAM05}.  Using the techniques 
described in \cite{RCL93}
we can estimated the Lyapunov exponent for this strange attractor: 
$l\approx 0.026$. A positive value of the Lyapunov exponent is another
indication of the problems facing the Horror scenario as past experience does
not provide detailed guidance on future behavior. In the next cycle the 
economy may
collapse more quickly or more slowly than it did in the previous cycle. This
type of behavior is typical of a dynamical system operating in the
chaotic regime \cite{CAM05}.

While previous research
has shown it is possible to avoid the chaos and collapse in the 
horror scenario by
setting $\kappa=100$ \cite{GWM96,MPF96,HL98}, such large values of
$\kappa$ seem unrealistic, because the first term in eq.~\ref{eq:freg} 
is $O(1)$ initially. 

Unfortunately, pollution abatement alone, does not move the system from
the phase of unsustainble development to the phase of nsustainble development, 
though it does
help to recover from an environmental collapse; in fact, as long
as the parameters remain within reasonable bounds they have no
impact on the phase diagram shown in Figures~\ref{fig:phase}.

\subsection{Pollution Control}

Pollution taxes have been introduced into the Wonderland model in
previous studies \cite{HL98,LH02,LPB03}.  The goal of a pollution tax is to
increase the effective decoupling rate, $\chi$, by making pollution
unprofitable.  In this paper, the pollution tax rate, $\tau$, enters the
model first via eq.~\ref{eq:p_state}, which changes to:

\begin{equation}
p(t+1)=p(t)\left(1-\chi-\chi_0\frac{\tau}{1+\tau}\right).
\label{eq:p_w_tax}
\end{equation}
where, $\chi_0$ is the maximum \textit{additional} decoupling rate for the 
assumed
level of technology, i.e., $\chi+\chi_0$ is the maximum achievable decoupling
rate for an assumed level of technology.

The side affect of taxing pollution is a reduction in the real per capita 
growth rate since resources are diverted into the pollution control; hence,
eq.~\ref{eq:y_state} becomes:

\begin{equation}
y(t+1)=y(t)\left(1+\gamma-\left(\gamma+\eta\right)\Big[1-z(t)\Big]^{\lambda}
             -\gamma_0\,\frac{\tau}{1-\tau}\right),
\label{eq:y_w_tax}
\end{equation}

\noi where the parameter $\gamma_0$ determines the amount by which the
pollution tax retards growth. Theoretically, a modest pollution tax does not 
diminish growth to the full extent of the tax, because the tax itself spurs
investment in pollution reduction technologies which in turn increase
growth; however, as the tax becomes larger, the numerator in
equation~\ref{eq:y_w_tax} tends toward zero and the tax can become a 
considerable drag on the economy.

By varying the decoupling rate, $\chi$ and the tax rate $\tau$, while keeping 
the other parameters fixed to their values in Table~\ref{tab:dream} (with
$\gamma_0=0.5$ and $\chi_0=\chi/2$), the phase diagram 
in Figure~\ref{fig:tax} can be constructed
using the techniques discussed in the last section. As can be seen, with the
help of a moderate pollution tax, the system can be moved from the phase of
unsustainable development to the phase of sustainable development.

In todays political climate many of society's leaders do not dispute the 
ability of pollution taxes or other remedies to improve the state of the 
world's natural capital, rather they
claim the cost in terms of foregone economic development is too high. Indeed,
many leaders would prefer to maintain high per capita growth now and deal with
the aftermath of an environmental collapse later, in the same manner the 
stock market bubble was allowed to expand at the end of 1990s until it burst
in 2000.

Concentrating on short term growth rates alone, however, can be misleading. As 
shown in Figure \ref{fig:volatility}, the average growth rate is the same
when the pollution tax rate is too low, or when the pollution tax is too high.
Hence, the average growth rate alone is not
definitive, policy makers must also be concerned about reducing economic 
volatility \cite{RR95,EIS01}, i.e., reducing fluctuations in the growth rate. 
Figure \ref{fig:volatility} also plots the variation in the normalized real
growth rate as a function of the pollution tax rate. As can be seen,
volatility drops to zero at the critical tax rate, which is also the point
where the real growth rate is a maximum. This is to be expected, since in
the phase of sustainable development, the Wonderland model contains no 
mechanism to create endogenous variability in the economic growth rate.

As an additional check on the viability of differing long term strategies, the
maximal regret\cite{SAV50} for alternative strategies should be calculated
and the strategy yielding the minimal of the maximal regret should be chosen.
Towards this end, we define the critical regret for per capita output for 
different value of the tax rate, $\tau$ as:

\begin{equation}
R_c(t,\tau)=\frac{y_c(t)-y(t,\tau)}{y_c(t)}
\label{eq:regret}
\end{equation}

\noi where $y_c(t)$ is the per capita output at the critical value 
of $\tau$, i.e., the value of $\tau$ on the phase boundary in
Figure~\ref{fig:tax} (all other parameters are assumed fixed). 
Figure~\ref{fig:regret} depicts the critical regret as a function of time 
for different
tax rates when all other parameters are held fix to the values they assume
in the Horror scenario. For tax rates lower than the critical tax rate, the
short term regret is negative, meaning faster economic growth than that
achievable at the critical tax rate; however, this faster growth is completely
dissipated once the environment collapses, leaving the long term regret at its
maximum possible value. For tax rates above the critical tax rate, economic
growth is slower, leading to increased regret. Hence, the critical tax rate is
the optimal tax rate in terms of Wonderland's long term prospects.

The relative ineffectiveness of pollution abatement versus pollution control
was noted previously by Leeves and Herbert \cite{LH02} who studied a modified
version of the Wonderland model in which eq.~\ref{eq:y_state} is replaced by a
Cobb-Douglas production function. Their model showed a transient dynamics
consisting of large volatility in the per capita
output and the stock of natural capital. As in the present model, pollution
abatement expenditures were ineffective at eliminating the unwanted behavior,
whereas adequate levels of pollution control expenditures were effective in
eliminating the volatility.

In summary, it seem reasonable to conclude that controlling pollution at the
source is more import as far as sustainability is concerned, than attempting
to abate its effects afterwards.

\section{Conclusion}

\begin{figure}[htbp]
\begin{center}
	\includegraphics[angle=90,width=\textwidth,keepaspectratio]{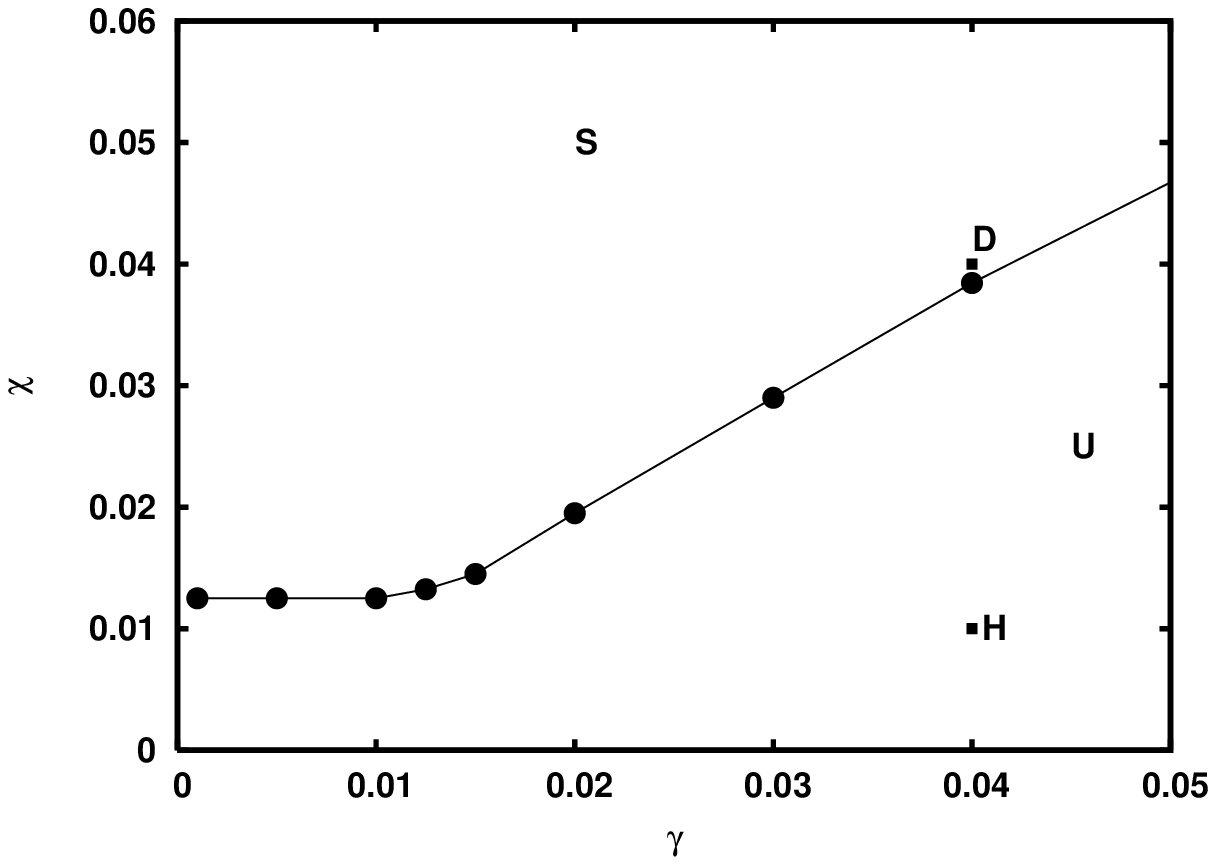}
    \caption{\label{fig:phase}}
\end{center}
\end{figure}

The advantages of basically treating eqs.~\ref{eq:x_state}-\ref{eq:f} as
dynamical equations of state have been demonstrated. The phase diagram
depicted in Figure~\ref{fig:phase} contains an infinite number of scenarios,
some of which have the property of sustainable development, others of which
do not. In the lower right hand corner, for example, one can find 
scenarios of environmental collapse due to over production; while in the
lower left hand corner are scenarios of collapse due to over population.

The analysis of the preceding two sections has yielded the information
needed to formulate an optimal strategy for ensuring the long term health of
Wonderland. Such a strategy would consist of the following steps: 

\begin{enumerate}

\item Determine the model parameters as accurately as possible.
\item Map out the phase diagram in that part of parameter space covered by
the parameters measured in step 1.
\item Determine the approximate critical pollution tax required for 
sustainable development
\item Periodically repeat steps 1-3 adjusting the pollution tax as necessary.

\end{enumerate}

\begin{figure}[htbp]
\begin{center}
	\includegraphics[angle=90,width=\textwidth,keepaspectratio]{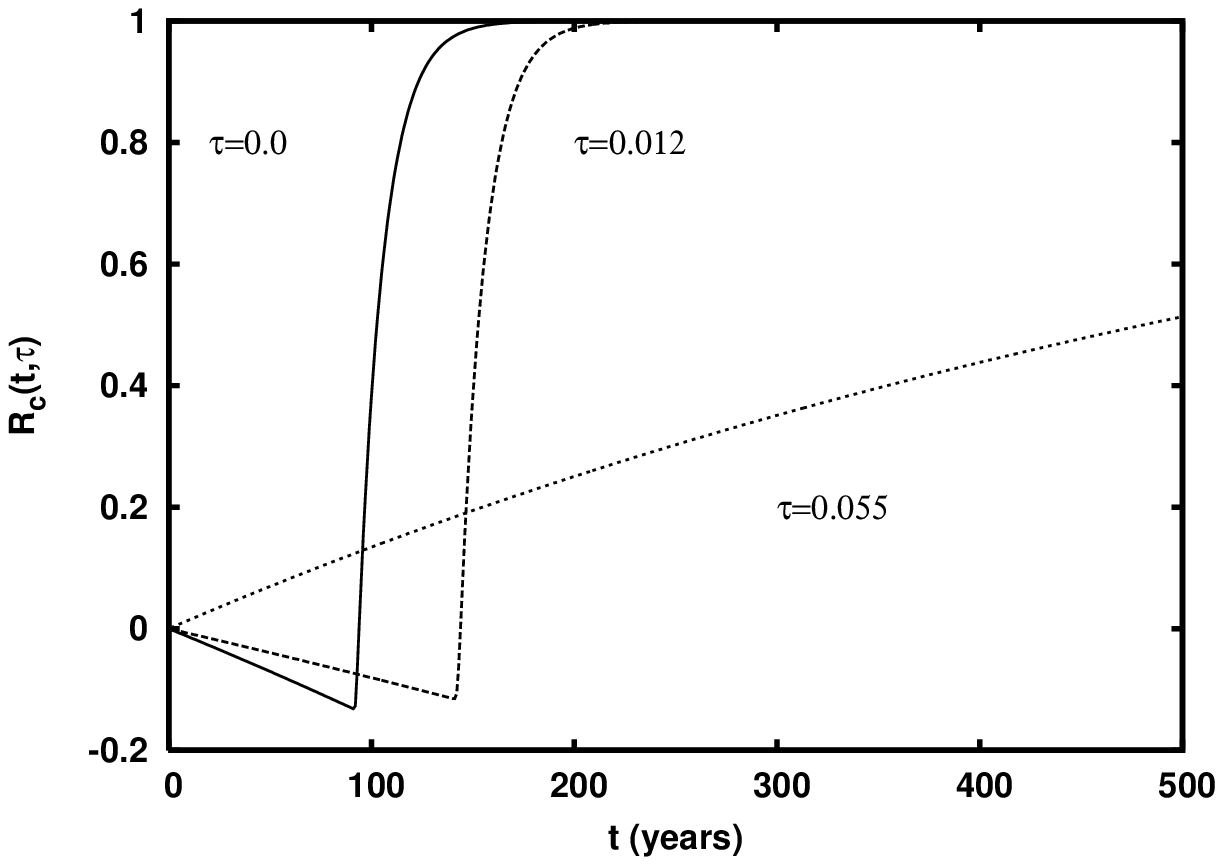}
    \caption{\label{fig:regret}}
\end{center}
\end{figure}

\noi As demonstrated by Figure~\ref{fig:regret}, this simple strategy is
optimal in the sense of yielding the minimal regret in the long term, at
the sacrifice of short term (and short lived) gains.

\begin{figure}[htbp]
\begin{center}
	\includegraphics[height=0.3\textheight,keepaspectratio]{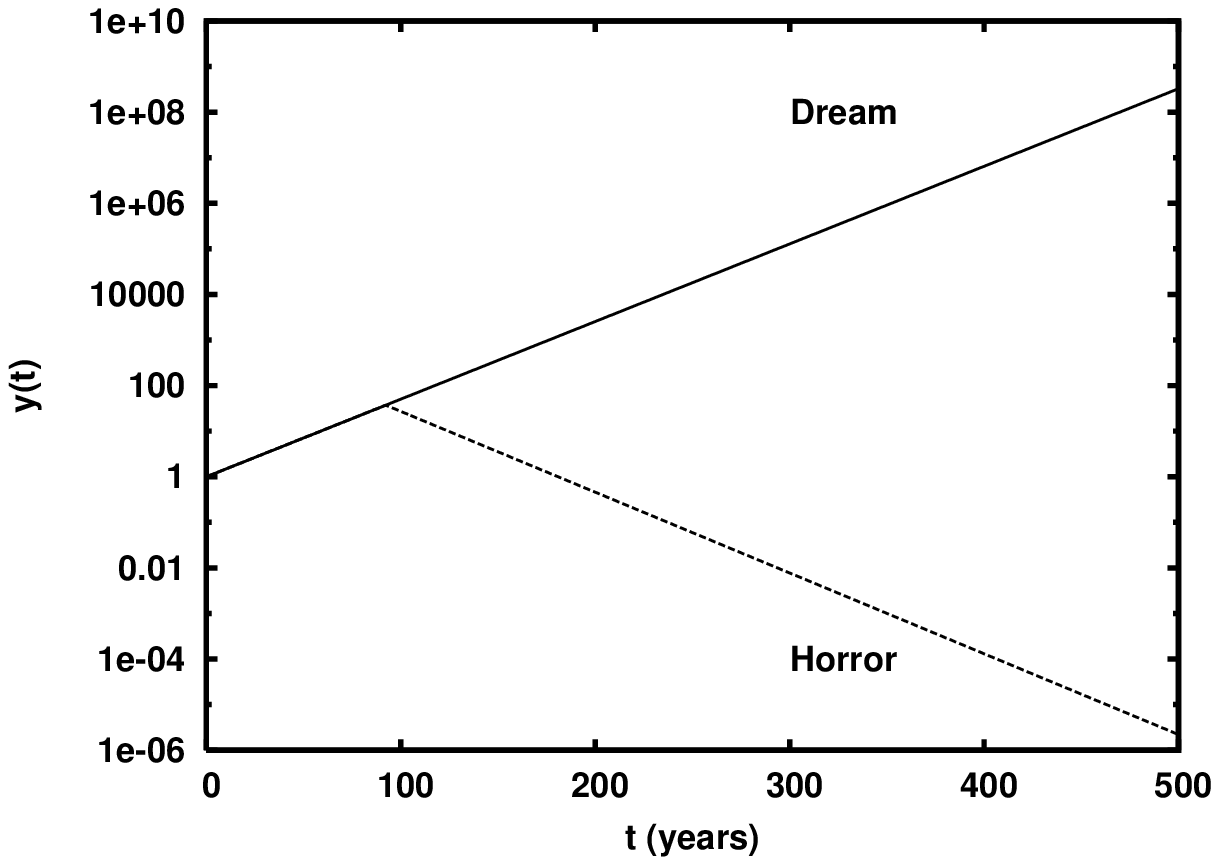}
    (a)
	\includegraphics[height=0.3\textheight,keepaspectratio]{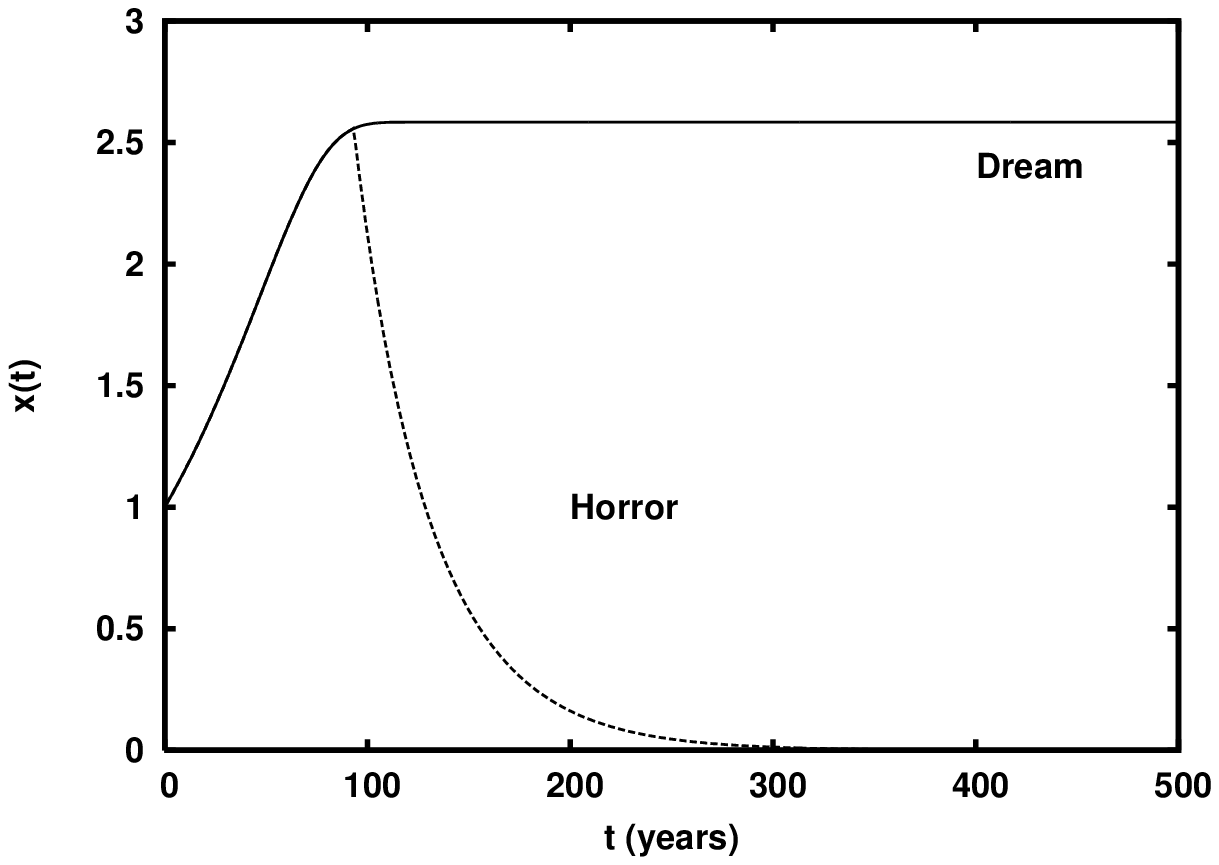}
    (b)
	\includegraphics[height=0.3\textheight,keepaspectratio]{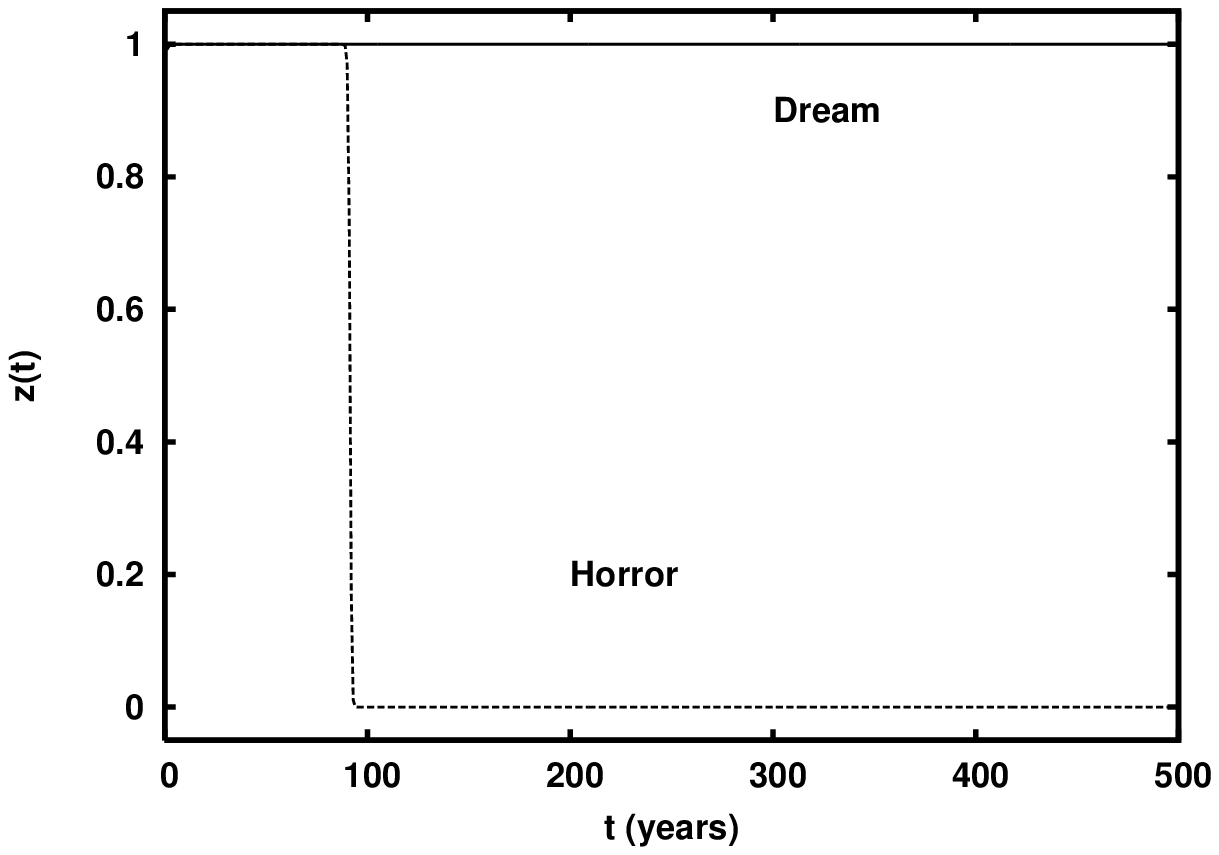}
    (c)
    \caption{\label{fig:traj}}
\end{center}
\end{figure}

\begin{figure}[htbp]
\begin{center}
	\includegraphics[angle=90,width=\textwidth,keepaspectratio]{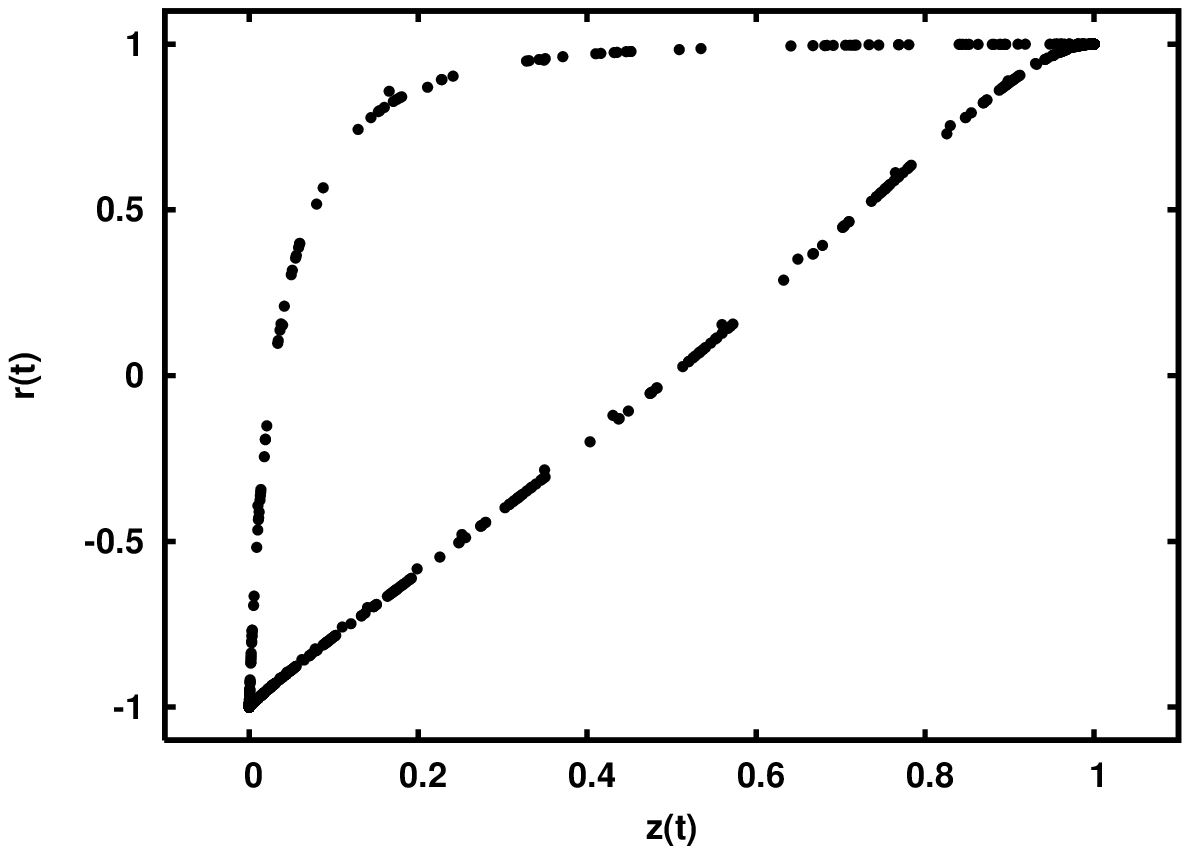}
    \caption{\label{fig:chaos}}
\end{center}
\end{figure}

One may question whether it is necessary to aim for the phase boundary, or
whether simply being ``not too far'' is sufficient, especially since the 
time to collapse increases exponentially as one approaches the phase boundary.
The answer is given by the discussion in \ref{sec:phase} and in particular
Figures \ref{fig:traj} and \ref{fig:chaos}; namely, the system provides 
little forewarning of an impending collapse. Relative stability may prevail for
decades, until the system suddenly undergoes a phase transition; hence, prudence
would dictate avoiding this parameter regime if at all possible.

As noted above, the Wonderland model is a comparatively simple integrated
model. For example, the maximum growth rate, $\gamma$, is
treated in this model as an exogenous parameter; in reality the
maximum growth rate is a complex function of the state of the economy, the
the environment and the population. When all three
collapse as in Figure \ref{fig:traj}, $\gamma$ would also be expected to fall,
leading to a faster collapse and longer recovery times. Hence, the
phase diagram in Figure \ref{fig:phase}
should be taken as upper bound on the sustainability due to the use of a
constant $\gamma$. An extension of the Wonderland model to include more 
realistic, endogenous growth through a Cobb-Douglas production function 
has been studied by Leeves and Herbert \cite{LH02}.

In summary, this paper has demonstrated how concepts originally developed for
studying critical phenomena in physical systems can be successfully applied to
problems in long-term, socio-economic planning. Of course, Wonderland is a
toy model which captures the global features of the complete
human-environment interactions, but not its details. Indeed, detailed models
transferable to the real world, where the models can contain an
order of magnitude more equations \cite{CHA00}; however, we believe the
principals elucidated here can still apply. Furthermore, the
principled approach developed here can augment the ensemble 
approach~\cite{LPB03} by applying large scale computer simulations more
systematically.

%\bigskip
%\bigskip
%\bigskip
%\noindent{\large\bf Acknowledgments}
%\smallskip
%
%\noindent The author would like to thank the reviewers for their useful and
%constructive commentary on this work.

\newpage

\bibliographystyle{unsrt}
\bibliography{sd}

\begin{thebibliography}{10}

\bibitem{CHA00}
R.~W. Chadwick.
\newblock Global modeling: Origins, assessment and alternative futures.
\newblock {\em Simulation \& Gaming}, 31:50--73, 2000.

\bibitem{MRM04}
D.~Meadows, J.~Randers, and D.~Meadows.
\newblock {\em Limits to Growth: The 30-Year Update}.
\newblock Chelsea Green Publishing Company, White River Junctions, Vermont,
  2004.

\bibitem{HUG99}
B.~B. Hughes.
\newblock {\em International Futures: Choices in the Search for a New World
  Order}.
\newblock Westview, Boulder, Colorado, 1999.
\newblock Available at:
  \href{http://www.du.edu/~bhughes/ifs.html}{http://www.du.edu/~bhughes/ifs.ht%
ml}.

\bibitem{RBG02}
P.~Raskin, T.~Banuri, G.~Gallop\'i n, P.~Gutman, A.~Hammond, R.~Kates, and
  R.~Swart.
\newblock {\em Great Transition: The Promise and Lure of the Times Ahead}.
\newblock Stockholm Environment Institute, Boston, Massachusetts, 2002.
\newblock Available at: \href{http://www.gsg.org/}{http://www.gsg.org/}.

\bibitem{CAM05}
P.~Cvitanovi\'c, R.~Artuso, R.~Mainieri, G.~Tanner, and G.~Vattay.
\newblock {\em Chaos: Classical and Quantum}.
\newblock Niels Bohr Institute, Copenhagen, Denmark, 2005.
\newblock Available at: \href{http://ChaosBook.org}{{http://ChaosBook.org}}.

\bibitem{BLP01}
S.~C. Bankes, R.~J. Lempert, and S.~W. Popper.
\newblock Computer-assisted reasoning.
\newblock {\em Computing in Science and Engineering}, 3:71--77, 2001.

\bibitem{LPB03}
R.~J. Lempert, S.~W. Popper, and S.~C. Bankes.
\newblock {\em Shaping the Next One Hundred Years: New Methods for
  Quantitative, Long-Term Policy Analysis}.
\newblock RAND, Santa Monica, California, 2003.
\newblock Available at:
  \href{http://www.rand.org/publications/MR/MR1626/}{http://www.rand.org/publi%
cations/MR/MR1626/}.

\bibitem{SAN94}
W.~C. Sanderson.
\newblock {\em Simulation Models of Demographic, Economic, and Environmental
  Interactions}, pages 33--71.
\newblock In Lutz \cite{LUT94}, 1994.

\bibitem{HL98}
R.~D. Herbert and G.~D. Leeves.
\newblock Troubles in wonderland.
\newblock {\em Complexity International}, 6:1--20, 1998.

\bibitem{LUT94}
W.~Lutz, editor.
\newblock {\em Population, Development, Environment: Understanding Their
  Interactions in Mauritius}.
\newblock Springer, Berlin, Germany, 1994.

\bibitem{COH95}
J.~E. Cohen.
\newblock {\em How Many People Can the Earth Support?}
\newblock W.W. Norton \& Company, Inc., New York, New York, 1995.

\bibitem{BLE05}
M.~Bar and O.~M. Leukhina.
\newblock A model of historical evolution of output and population.
\newblock Technical report, Department of Economics, University of Minnesota,
  July 2005.
\newblock Available at:
  \href{http://www.stanford.edu/group/SITE/papers2005/Leukhina.05.pdf}{http://%
www.stanford.edu/group/SITE/papers2005/Leukhina.05.pdf}.

\bibitem{EH71}
P.~R. Ehrlich and J.~Holdren.
\newblock Impact of population growth.
\newblock {\em Science}, 171:1212--1217, 1971.

\bibitem{GWM96}
E.~Gr\"oller, R.~Wegenkittl, A.~Milik, A.~Prskawetz, G.~Feichtinger, and W.~C.
  Sanderson.
\newblock The geometry of wonderland.
\newblock {\em Chaos, Solutions \& Fractals}, 7:1989--2006, 1996.

\bibitem{MPF96}
A.~Milik, A.~Prskawetz, G.~Feichtinger, and W.~C. Sanderson.
\newblock Slow-fast dynamics in wonderland.
\newblock {\em Environmental Modeling and Assessment}, 1:3--17, 1996.

\bibitem{LH02}
G.~D. Leeves and R.~D. Herbert.
\newblock Economic and environmental impacts of pollution control in a system
  of environment and economic interdependence.
\newblock {\em Chaos, Solitons \& Fractals}, 13:693--700, 2002.

\bibitem{SMA76}
S.-K. Ma.
\newblock {\em Modern Theory of Critical Phenomena}.
\newblock Addison-Wesley, Reading, Massachusetts, 1976.

\bibitem{RCL93}
M.~T. Rosenstein, J.~J. Collins, and C.~J.~De Luca.
\newblock A practical method for calculating largest lyapunov exponents from
  small data sets.
\newblock {\em Physica D}, 65:117--134, 1993.

\bibitem{RR95}
V.~A.~Ramey G.~Ramey.
\newblock Cross-country evidence on the link between volatility and growth.
\newblock {\em American Economic Review}, 85:1138--1151, 1995.

\bibitem{EIS01}
W.~Easterly, R.~Islam, and J.~E. Stiglitz.
\newblock Shaken and stirred: Explaining growth volatility.
\newblock Technical report, The World Bank, January 2000.
\newblock Available at:
  \href{http://www.worldbank.org/research/abcde/washington\_12/pdf_files/easte%
rly.pdf}{http://www.worldbank.org/research/abcde/washington\_12/pdf\_files/eas%
terly.pdf}.

\bibitem{SAV50}
L.~J. Savage.
\newblock {\em The Foundations of Statistics}.
\newblock John Wiley \& Sons, New York, New York, 1950.

\end{thebibliography}

\newpage
\appendix{\large\bf Tables}

\bigskip
\bigskip

\begin{table}[htbp]
\begin{center}
\begin{tabular}{||m{1cm}|m{1cm}||m{1cm}|m{1cm}||m{1cm}|m{1cm}||}
\hline
\multicolumn{2}{||c||}{\textbf{Economy}} &
\multicolumn{2}{c||}{\textbf{Environment}} &
\multicolumn{2}{c||}{\textbf{Population}}\\
\hline
name & value & name & value & name & value \\
\hline
$\gamma$ & 0.04 & $\chi$ & 0.04 & $\alpha$ & 0.09 \\
\hline
$\eta$ & 0.04 & $\delta$ & 1.0 & $\alpha_0$ & 10.0 \\
\hline
$\lambda$ & 2.0 & $\rho$ & 0.2 & $\alpha_1$ & 2.5 \\
\hline
& & $\omega$ & 0.1 & $\alpha_2$ & 2.0 \\
\hline
& & & & $\beta$ & 0.8 \\
\hline
& & & & $\beta_0$ & 40.0 \\
\hline
& & & & $\beta_1$ & 1.375 \\
\hline
& & & & $\theta$ & 15.0 \\
\hline
\end{tabular}
\caption{\label{tab:dream}Parameter values for the Dream scenario.}
\end{center}
\end{table}

\newpage
\noi\appendix{\large\bf Figures}
\bigskip

\bigskip
\noi Figure \ref{fig:traj} The basic scenarios Dream and Horror. (a) Semi-log 
plot of the per capita output as a function of time. (b) Plot of relative
population as a function of time. (c) Plot of natural capital as a function of
time.

\begin{figure}[htbp]
\begin{center}
	\includegraphics[angle=90,width=\textwidth,keepaspectratio]{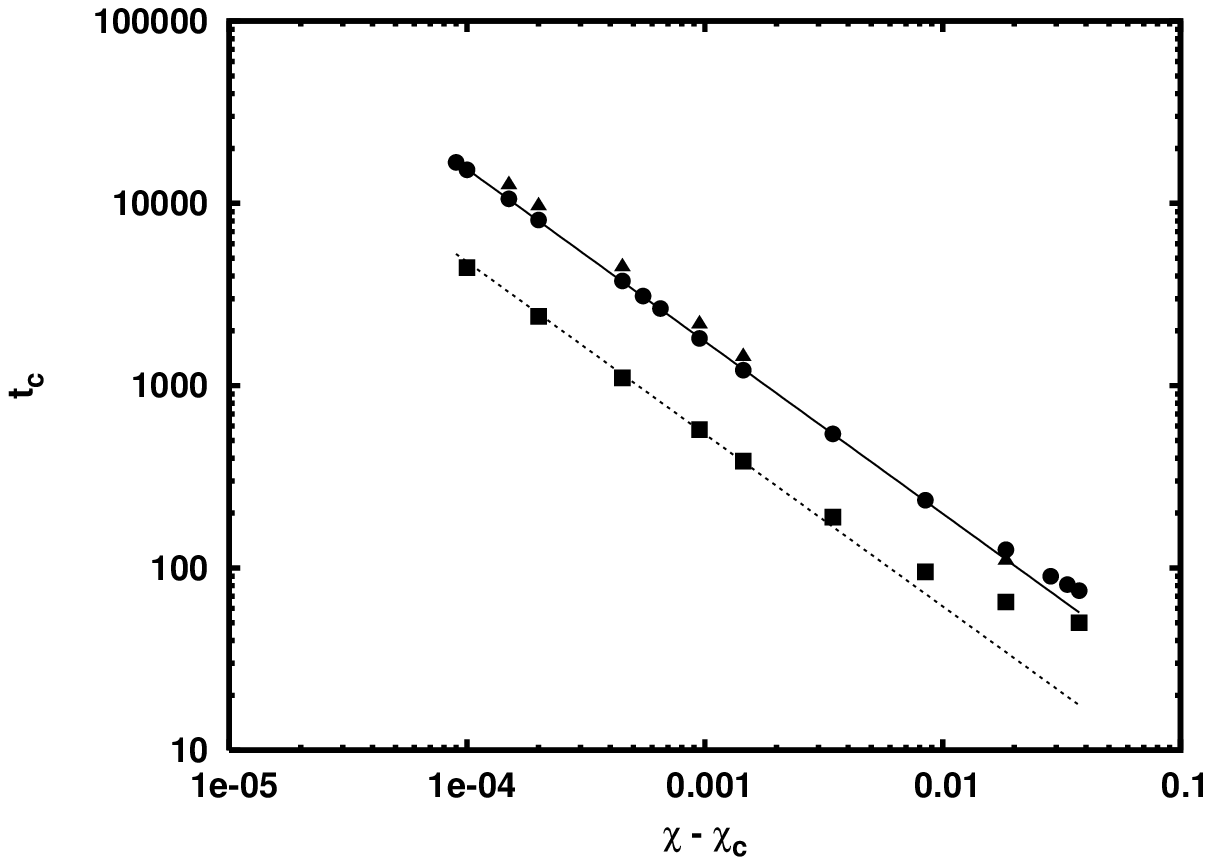}
    \caption{\label{fig:scale}}
\end{center}
\end{figure}

\bigskip
\noi Figure \ref{fig:scale} Log-log plot of the time before the first 
collapse of natural capital as a function of the distance $\chi-\chi_c$.
Different point types indicate different initial states of the system.

\bigskip
\noi Figure \ref{fig:phase} Phase diagram in the $\gamma-\chi$ plane. The 
line marks the boundary between the phases of sustainable, S,
and unsustainable, U, development. H marks the location of the Horror
scenario, while D marks the location of the Dream scenario.

\bigskip
\noi Figure \ref{fig:chaos} Plot of the normalized, real growth rate as a 
function of the stock of natural capital over a $40\, 000$ year time span.

\begin{figure}[htbp]
\begin{center}
	\includegraphics[angle=90,width=\textwidth,keepaspectratio]{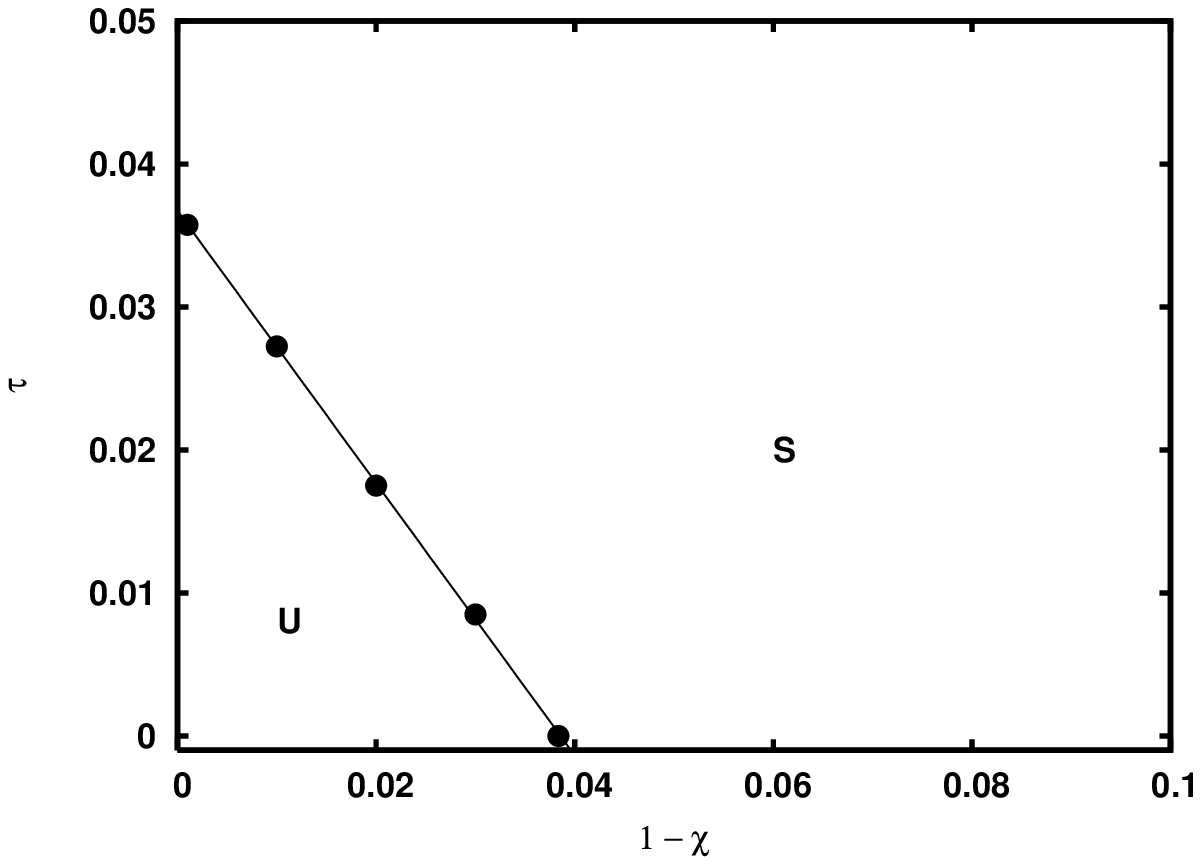}
    \caption{\label{fig:tax}}
\end{center}
\end{figure}

\bigskip
\noi Figure \ref{fig:tax} Phase diagram in the $\tau-\chi$ plane
when the other parameters are fixed to their values in Table~\ref{tab:dream} 
and $\gamma_0=0.5$.

\begin{figure}[htbp]
\begin{center}
	\includegraphics[angle=90,width=\textwidth,keepaspectratio]{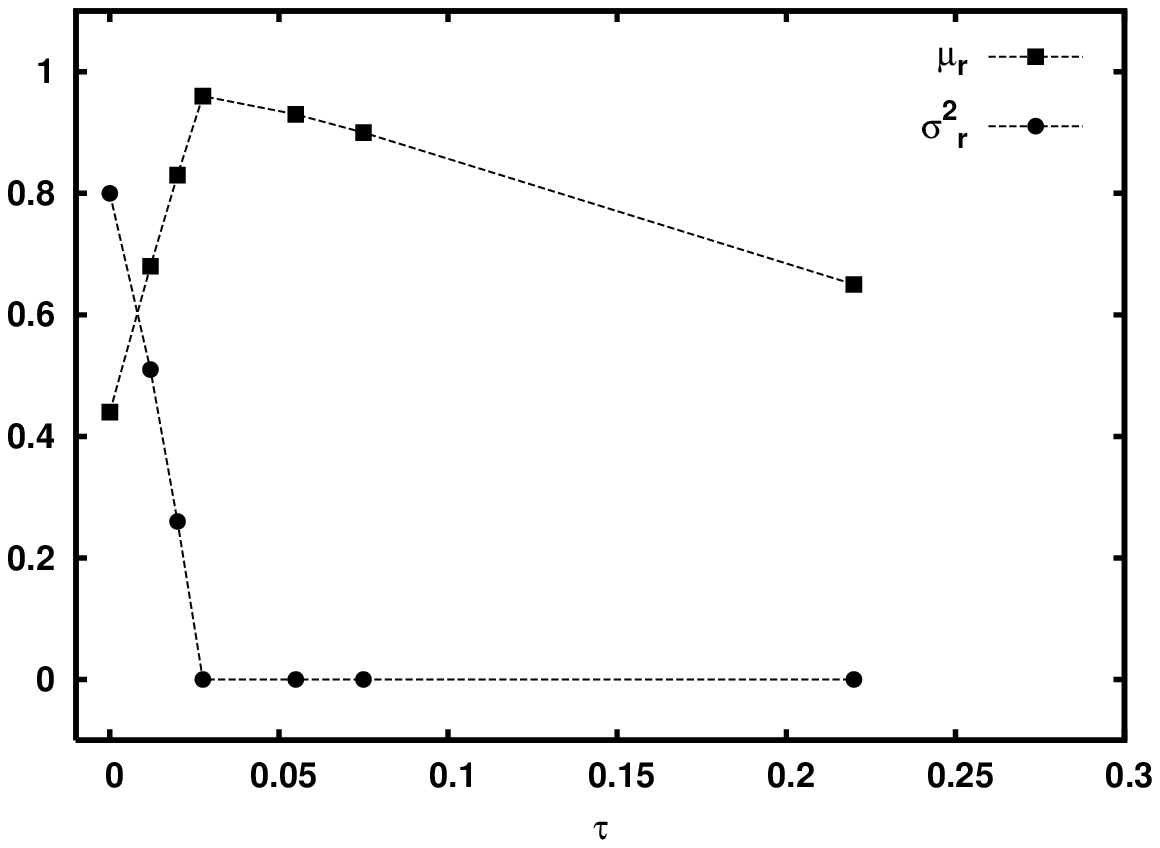}
    \caption{\label{fig:volatility}}
\end{center}
\end{figure}

\bigskip
\noi Figure \ref{fig:volatility} The average, $\mu_r$, and deviation,
$\sigma^2_r$ of the normalized real growth rate, $r(t)$ (see eq. 
\ref{eq:r}) as a function of the tax rate, $\tau$, when the other parameters 
are the same as those used in the Horror scenario.

\bigskip
\noi Figure \ref{fig:regret} The critical regret as a function of time
for different values of the tax rate, $\tau$, when the other parameters are 
the same as those used in the Horror scenario.

\bigskip

\newpage

\newpage

\newpage

\newpage

\newpage

\newpage

\newpage

\end{document}